# Continuous Analysis: Evolution of Software Engineering and Reproducibility for Science

Venkat S. Malladi[1], Maria Yazykova[2], Olesya Melnichenko[3] and Yulia Dubinina[4]

Health Futures,
Microsoft, Redmond WA 98052, USA

**Abstract**

Reproducibility in research remains hindered by complex systems involving data, models, tools, and algorithms. Studies highlight a reproducibility crisis due to a lack of standardized reporting, code and data sharing, and rigorous evaluation. This paper introduces the concept of Continuous Analysis to address the reproducibility challenges in scientific research, extending the DevOps lifecycle. Continuous Analysis proposes solutions through version control, analysis orchestration, and feedback mechanisms, enhancing the reliability of scientific results. By adopting Continuous Analysis, the scientific community can ensure the validity and generalizability of research outcomes, fostering transparency and collaboration and ultimately advancing the field.

## 1  Introduction

Reproducibility is a cornerstone of both scientific research and software development. Reproducibility in computational sciences extends beyond simply making sure that the original researcher can replicate their results. It requires that researchers or organizations can achieve the same outcomes using shared data, code, and methods.

In the open-source community, particularly in large projects like the Linux Kernel, reproducibility is achieved by adopting practices that enable rapid development and collaboration from a diverse group of contributors. This is done using continuous processes that facilitate the integration and deployment of code, which ensure that contributions are tested, integrated, and released in a reliable manner. These continuous processes, Continuous Integration (CI) and Continuous Deployment (CD), are critical components of DevOps methodology that emerged in late 2000 – early 2010. In the context of software engineering, DevOps lifecycle has become standard practice, promoting consistent, automated testing and deployment. Its procedures ensure that code contributions from multiple developers can be integrated smoothly while maintaining the reliability of the final product (Kim et al., 2016).

Later several extensions to DevOps methodology were proposed: DataOps and MLOps.  DataOps is a set of collaborative practices to manage the entire data life cycle from data ingestion and transformation to validation and delivery. Its primary goal is to provide maximum value from the data (Ereth, 2018; Mainali et al., 2021; Mucci et al., 2024). MLOps is a methodology that encompasses best practices, key concepts, and a development culture for the entire lifecycle of machine learning products. It involves end-to-end processes, including conceptualization, implementation, monitoring, deployment, and scalability. Its goal is to bring machine learning systems into production (Lwakatare et al., 2020; Ritz et al., 2022; Kruzeberger et al., 2023).

[1] vmalladi@microsoft.com – corresponding author
[2] mariayaz@microsoft.com
[3] olesya.melnichenko@microsoft.com
[4] yuliadub@microsoft.com

However, in scientific research and rapidly evolving technologies, reproducibility remains a challenge due to the increasing complexity of systems that now must include data, models, tools, and algorithms. To address this, we propose extending the DevOps lifecycle to include Continuous Analysis(CA), a concept introduced by Beaulieu-Jones and Greene (2017) to promote reproducibility in science. Continuous Analysis functions as an overarching layer that not only integrates development, testing, and deployment lifecycle but also addresses the gaps and shortcomings inherent in CI/CD processes, ensuring that results can be consistently and reliably reproduced as algorithms, models, and datasets evolve over time.

By incorporating continuous analysis into the research practice, we can ensure that scientific workflows are not only efficient but also reproducible, facilitating greater transparency and reliability in research and development. This approach fully encapsulates the lifecycle of scientific work, from initial development to monitoring of the final deployment, ensuring that every step can be traced, tested, and reproduced by anybody.

## 2     The current state of development and reproducibility in science

One of the main challenges facing science development, such as primary analysis, AI and Large Language Model (LLM), is ensuring the reproducibility and robustness of their results. Reproducibility denotes the capability to achieve consistent results, within acceptable tolerances, when applying the same methods and data despite inherent variability in algorithms. Robustness, on the other hand, signifies the ability to sustain the performance and validity of methods and data across varying conditions and scenarios. Both aspects are crucial for ensuring the reliability, trustworthiness and accountability of analysis systems, especially in domains where they have significant social and ethical implications, such as health care, education, or justice.

Several studies have shown that there is a lack of reproducibility and robustness in many AI and LLM publications and applications. A survey by Hutson (2018) found that only 15% of 400 AI papers published in 2018 shared their code, and only 30% shared their data. Another replication study by Raffel et al. (2019) found that only 55% of 255 natural language processing (NLP) papers published in 2017 and 2018 provided enough information and resources to reproduce their results, and only 34% of the reproduced results matched or exceeded the original ones. Moreover, a benchmarking study by Belz et al. (2021) found that only 14.03% of 513 original/reproduction score pairs matched, demonstrating the difficulty of replicability in NLP. Additionally, 59.2% of reproduction results under the same conditions were worse, highlighting the reproducibility crisis and the need for reliable, stable, and generalizable NLP outcomes. Similarly, there is a reproducibility crisis across the sciences.

In computational biology, reproducibility problems often arise from complex workflows, software dependencies, and data sharing practices. A review of 15,000 bioinformatics tools published in over 2,000 studies showed that nearly 50% of the tools were difficult to install or reproduce due to poorly documented dependencies and the complexity of biological datasets (Mangul et al., 2019). Additionally, in computational fields like physics and chemistry, reproducibility has become a significant challenge, particularly as experiments rely more on complex software and large datasets.

In a study focused on computational materials science, researchers highlighted four key issues affecting reproducibility: lack of reporting computational dependencies, missing version logs, poorly organized code, and unclear references in manuscripts. These issues create substantial barriers for others trying to reproduce the results of machine learning-based predictions in materials research (Persaud et al, 2023).

These findings suggest there is a need to improve the development lifecycle practices used in research and industry to address reproducibility. Some of the factors that contribute to the reproducibility and robustness issues include the lack of standardized and transparent reporting of methods and results, the lack of sharing and accessibility of code and data, the lack of rigorous and comprehensive evaluation and testing of models, and the lack of awareness and incentives for adopting best practices and tools. Therefore, we propose that continuous analysis, as an extension of continuous integration and continuous deployment, can help address these challenges and foster a culture of reproducibility and robustness in scientific development.

## 3 What is Continuous Analysis?

Continuous analysis (CA) is a process that extends the principles and tools of continuous integration and continuous deployment to the analysis of data, code and models together, ensuring that they are always up-to-date, consistent, validated, and reproducible. CA is built on the principles that lay in the base of DevOps methodology. We begin by outlining these principles, which we then break down into distinct sections. Later we describe general CA workflow.

**Principles of Continuous Analysis**

- **Version Control: Tracking changes for code, software dependencies and data**
- **Analysis orchestration: Continuous Integration, Continuous Delivery, Continuous Monitoring**
- **Feedback: automated testing, artifact collection and storage, metrics, quality assessments, performance monitoring, evaluations, collaboration and documentation**

By adopting continuous analysis, researchers can benefit from an automated workflows that facilitate documentation, sharing, testing, and deployment of their code and data, as well as the generation and dissemination of their results, facilitating a more "open science". Continuous analysis can also enable more collaborative and interdisciplinary research and development, as it allows for easier and faster replication, comparison, and integration of different methods and datasets across different domains and tasks.

### 3.1 Version control

Version control is essential for tracking changes in data, code, and software dependencies. It ensures that every modification is recorded, allowing researchers to revert to previous versions if necessary. This principle is fundamental for maintaining consistency, accountability, and collaboration among team members working on a project.

**Code** version control is the practice of tracking and managing changes to software code. Version control systems (git, Azure DevOps, GitLab, etc.) help teams manage changes to the code over time while many people edit the code, help with continuous deployment and tracking of issues. It also allows roll back capability to working version of code if issues with newer features are discovered.

Code version control is especially important in research settings since so many iterations and adjustments are made over short periods of time. If there is no record of each of those changes, that greatly impacts the reproducibility of research results. It is not just version control that is important, but also proper documentation that comes with adding changes. This means having description commit messages, as well as helpful Pull Request descriptions when it comes to documenting the ongoing changes to the code.

**Software dependencies** It is important to have a consistent way to outline and manage software dependencies for any piece of code. For example, in python, dependencies can be managed via a requirments.txt file or using a tool like Poetry. It is also important that the dependencies file is properly tracked under code version control to ensure an easy way to see if an update to it has created issues in the subsequent deployment. In the context of scientific research, once you have code and dependencies that are tracked under code version control, the utilization of containers (i.e. Docker) offers a reproducible and consistent environment for computational experiments.

**Data** version control is important for scientific analysis, as it enables researchers to ensure the reproducibility, traceability, and quality of their data and results. By keeping track of data provenance, transformations, and dependencies, data version control can help researchers avoid data loss, corruption, inconsistency, and duplication.

One of the main challenges of data version control is keeping the data and the code that operates on it connected and synchronized, as they may change independently and frequently. This can lead to inconsistencies, errors, and conflicts in the analysis and results, and difficulties in tracking and reproducing the data and code versions. To address this challenge, data version control tools and frameworks need to provide mechanisms to link the data and the code, such as metadata, checksums, hashes, tags, branches, and workflows, and automate the synchronization and update of the data and the code whenever they change.

### 3.2 Feedback

Feedback is a critical component of Continuous Analysis, providing valuable insights that can inform and improve the analysis and development process. Continuous Analysis features multiple feedback loops that allow the developers multiple opportunities for inspection of the whole process (Figure 1). This includes traditional feedback loops of CI/CD. As new code or data is integrated and deployed, automated tests and performance benchmarks can provide real-time feedback to developers on issues such as failing tests, performance regressions, or security vulnerabilities. This allows for rapid correction and refinement of models, algorithms, or applications.

Additionally, Continuous Monitoring of quality metrics, task performance, and evaluations are included. For instance, quality metrics can include accuracy, precision, recall, and other relevant performance indicators that help in evaluating the effectiveness of the models and analyses being conducted. Monitoring task performance ensures that each step in the analysis pipeline is executed efficiently and correctly, minimizing errors and delays. Evaluations can involve manual peer reviews, automated testing, and validation against known benchmarks or datasets. By regularly evaluating the results, teams can gain a deeper understanding of the strengths and weaknesses of their methods, leading to more informed decisions about future improvements and adjustments.

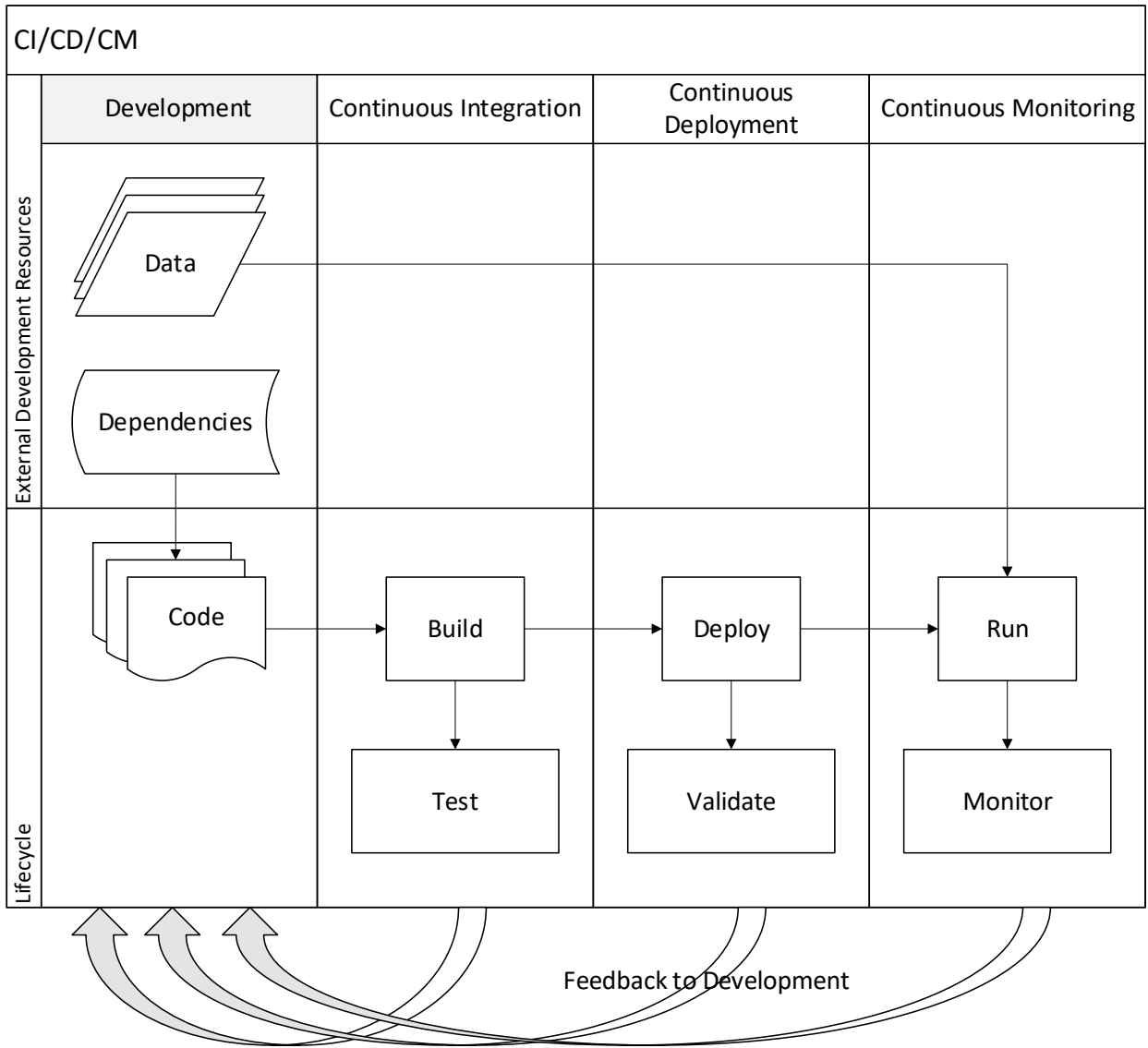

**Figure 1. Continuous Integration/Continuous Deployment/Continuous Monitoring Overview:** Code and Data are the parts of the Development that need to be versioned. Dependencies are tracked in code and represent external entities and relations. Continuous Integration, Continuous Deployment, and Continuous Monitoring, together represent CI/CD/CM pipeline that encapsulates a process of release creation. Automation is crucial in CI/CD/CM pipeline, as it ensures extensibility, reproducibility and supports agile development. Feedback is a collection of artifacts/signals containing items like logs, input parameters and commands, outputs, telemetry, and captured system states, gathered at each stage of CI/CD/CM pipeline. The feedback is analyzed and used for recommendations to be implemented in code (or data). Data is implicitly related to code (data schema, storage location, type of data, etc., need to be accounted in design, implementation, and tests), but the most impact is delayed until code runs on real data vs a test dataset. And when real data is used/processed, monitoring for data-related signals and issues is essential, in addition to other functional metrics.

For each feedback loop, continuous analysis encourages collecting and storing artifacts that are tagged to particular code/data changes. For code development, artifacts can be compiled binaries, Docker images, or test results. In data development, artifacts can be datasets, pre-processed data, or model checkpoints, which are saved during experimentation. Artifacts allow developers to ensure that each code or data change is traceable and reviewable. Tools like GitHub Actions enable this by automatically archiving build outputs and making them accessible for future analysis or peer review.

These artifacts, whether related to code or data, create a traceable, documented history that encourages collaboration, experimentation, transparent peer reviews, and informed decision-making throughout the development cycle. These artifacts also provide checkpoints during development to be able to manually interrogate the progress as well as providing a timeline for released versions.

### 3.3  Analysis orchestration

Automation plays a pivotal role in enhancing reproducibility by ensuring that processes are consistent, repeatable, and less prone to human error.

**Continuous integration (CI)** involves automatically building and testing the logic of various parts of the repository for the expected behavior. Thus, ensuring that any changes that would unexpectedly alter the outcome or cause undue behavior, are caught early and fixed. CI tools and systems, such as Jenkins, GitLab CI, Travis CI, and Azure DevOps, use automated approaches to streamline the development process by continuously merging code changes, running tests, and ensuring the codebase remains functional and deployable, enabling faster delivery cycles and higher code quality.

**Continuous deployment (CD)** focuses on how to automatically deploy acceptable changes to a production-like environment. This ensures code changes are delivered to users seamlessly while also receiving immediate feedback for users on changes. CD tools and systems, such as CircleCI, GitHub Actions, Spinnaker, and AWS CodeDeploy, automate the release process by automatically deploying every code change that passes automated tests, ensuring that new features and fixes are delivered to users quickly and reliably, while minimizing manual intervention and reducing the risk of deployment errors.

**Continuous Monitoring** focuses on coordinating the complex graph of tasks based on dependencies and execution order, ensuring that each task is executed in the correct sequence and at the appropriate time. Automated workflows guarantee that analysis tasks are consistently performed in the same manner every time. This uniformity eliminates the variability often observed in manual processes, where individuals might follow slightly varying procedures. By automating tasks such as data pre-processing, model training, and evaluation, teams can ensure adherence to consistent steps for each analysis run, leading to more reliable and reproducible results.

For example, automated workflows in Azure Synapse manage complex ETL processes by integrating data from various sources, processing it with Apache Spark or SQL, and loading it into data warehouses. These workflows handle large-scale data transformations using connectors for Azure Data Lake, Blob Storage, and external systems. In Azure Machine Learning, workflows manage tasks from data preparation to model deployment, allowing for efficient scaling, reuse, and version tracking. For environment-agnostic workflow orchestration engines, you can use Apache Airflow (Apache Airflow Documentation) or Nextflow (Di Tommaso et al. 2017), which help manage and automate complex data processing and scientific workflows. Airflow focuses on batch processing using Directed Acyclic Graphs (DAGs), while Nextflow excels in scalable, portable workflows using a dataflow programming model.

Overall, automation schedules and execution of tasks based on specific events, ensure timely data processing, reducing human error, and enabling real-time insights. It also scales efficiently to manage large systems and complex dependencies, supporting concurrent development and experimentation.

### 3.4 Continuous analysis workflow

There are six main types of artifacts to keep track of: data, code, code dependencies, tests, deployment artifacts (e.g. configurations and images), and results (figures, data, metrics, etc.). Each artifact type has a dedicated storage system (e.g. code repository) that may be linked, explicitly or implicitly, to a big graph. The code dependencies are references to other graphs of external development artifacts. Similarly, data artifacts are governed by data management systems and have relations to other datasets via data lineage. Although the data is not explicitly linked to code, the data processing code that runs during an experiment is written with assumptions about the data format and tools deployed in the processing environment. Therefore, any meaningful connections between artifacts need to be captured and covered by automation that is triggered by changes or events in each artifact repository.

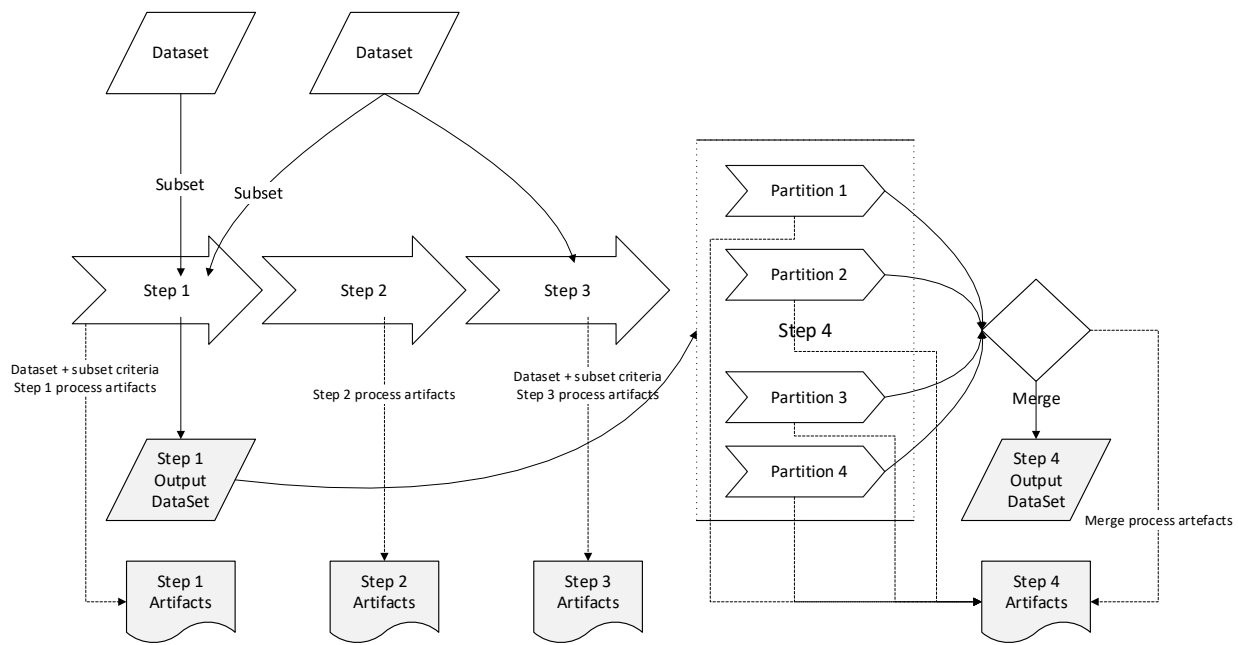

**Figure 2. An example of an experiment run graph:** The artifact boxes and the output datasets for each step are shaded. The outcome dataset created by Step 1 feeds into Step 4 input, while Step 4 is a partitioned process and includes merging logic for outcomes.

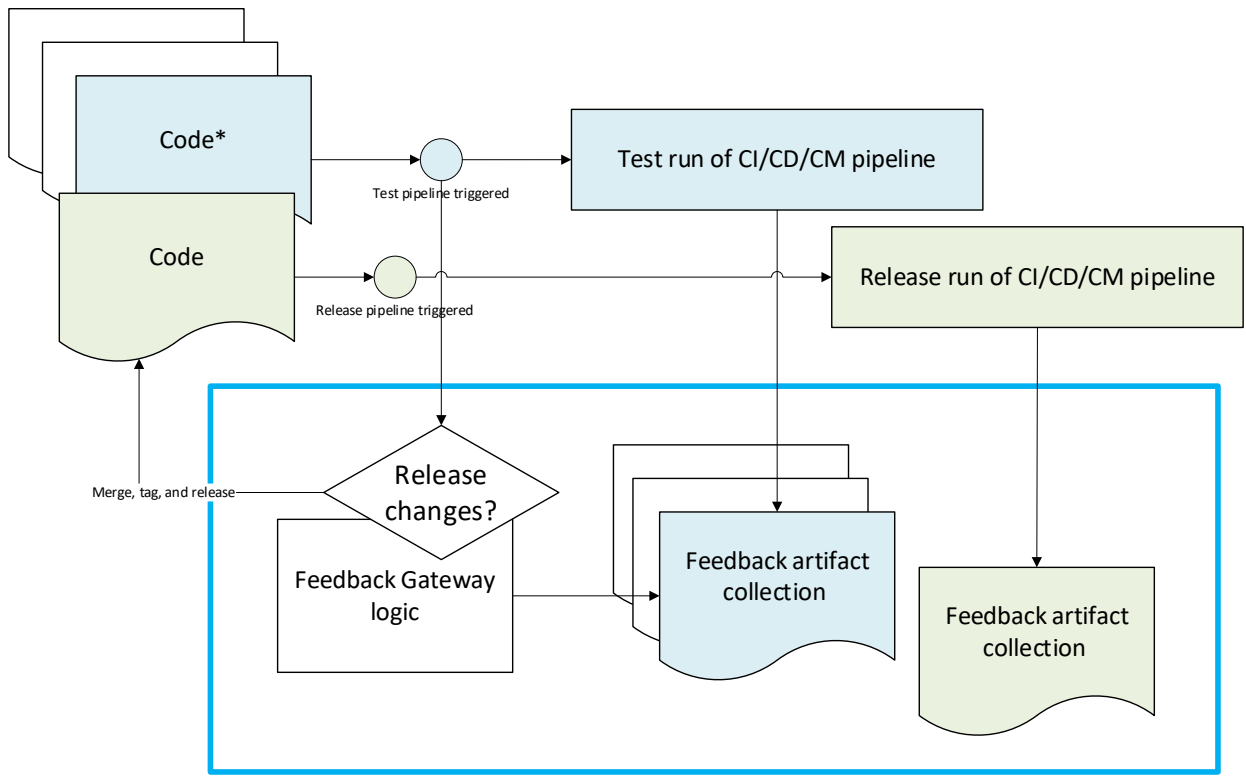

**Figure 3. Continuous Analysis workflow overview:** Code* represents modified code in working code branch. The blue shaded boxes indicate processes in the working branch, while the green shaded boxes represent processes in the release/main branch. The white box outlined by blue border encapsulates additional feedback artifact storage and Feedback Gateway logic added for Continuous Analysis.

As Figure 1 illustrates, an actual experiment run consists of a specific version of each component, such as code, dependencies, deployment, and data – collectively referred to as the artifact version tuple. It is essential to link the outcomes of the experiment run to this tuple, and the experiment run itself should be assigned a unique identifier. To enable Continuous Analysis, each run of experiment, test or production, needs to be instrumented to save run identifier with artifact version tuple and results (outputs). Consider an example of an experiment run illustrated in Figure 2. Each step of a flow, input resources and outcomes with artifacts can be presented as a graph. The flow graph analysis is helpful to identify inputs and outcomes of each step and categorize artifacts by their impact on the experiment outcome. Once critical artifacts are identified, they can be added to the artifact version tuple.

A diagram on Figure 3 illustrates the concept of continuous analysis by depicting the various stages and components involved in the process. When new data is added to the data repository, or when data dependencies are updated, an event is issued to the code repository to initiate the validation pipeline on the continuous integration, deployment, and feedback (CI/CD/CM) for the new version of data and the current versions of other repositories. The results from these validation runs are subsequently stored in the results repository, where they can be accessed and visualized by developers or other stakeholders. Upon successful validation and approval of the results, the data version and references to the results are updated in the main branch of the code repository, triggering the release pipeline on the main branch. The identical procedure is applied when there are modifications to the code or dependencies.

Notice that we suggest test pipeline (highlighted in blue in Figure 3) to include at least one deployment and a test experiment run using a subset of external (to code repository) data. The results of this test run, along with the artifact version tuple, should be added to the results storage alongside branch and change information. The release pipeline (highlighted in green) includes a full set of experiment runs, either automatically or manually performed, maintaining the same level of artifact granularity as gathered during the test pipeline. This requirement ensures that the tuples of feedback artifact collection are aligned and thus comparable.

## 4    Discussion

Continuous Analysis (CA) brings undeniable benefits to computational research by improving reproducibility, code quality, and collaboration. However, it's critical to recognize the challenges associated with it.

**Complexity and Resource Overhead**

Various "Ops" like DevOps, DataOps, and MLOps each address different aspects of computational and data-driven research, focusing on code, data, or models. These principles share common goals but require a comprehensive end-to-end approach for true reproducibility. Setting up a continuous analysis pipeline can be technically complex, requiring significant initial investment in infrastructure and configuration. The need to integrate tools like Docker, Jenkins, Azure Batch, and GitLab CI/CD can overwhelm teams unfamiliar with these technologies.

The costs associated with cloud computing resources (e.g., for storage and compute instances) and maintenance of the pipeline infrastructure can be high. Depending on the scale and needs of the project, sustained funding might be necessary to support long-term use of CA. Research teams must evaluate whether the return on investment in terms of efficiency and reproducibility justifies these upfront and ongoing costs.

**Increased Development Time**

In the initial stages of a project, implementing CA might extend the development timeline due to the need for thorough configuration and testing of pipelines. Unlike traditional methods, where teams typically focus on getting research outputs quickly, the CA approach requires a slower initial setup phase. This could discourage researchers, especially in fast-paced fields, where immediate results are often prioritized. However, once CA pipelines are established, they can save time by automating repetitive tasks and reducing errors, so it becomes a matter of balancing the short-term cost against long-term efficiency gains.

**Cultural and Organizational Changes**

Adopting CA is not just a technical shift but also a cultural transformation. For research teams to fully benefit from CA, they must develop a culture that prioritizes reproducibility, transparency, and collaboration. Encouraging researchers to adopt CA practices may require organizational support, such as incentives for publishing reproducible code and data. Research institutions and funding bodies can play a pivotal role by providing infrastructure, training, and incentives to promote CA adoption. For example, institutions might require reproducibility audits as part of the peer-review process or offer awards for best practices in reproducibility.

In conclusion, while Continuous Analysis offers considerable advantages in terms of reproducibility and research efficiency, it also introduces challenges that must be addressed. Research teams need to carefully consider the initial resource investment, development time, and training requirements, while fostering an organizational culture that values reproducibility. By addressing these challenges strategically, CA can ultimately streamline research workflows and contribute to more reliable, impactful scientific outcomes.

## Acknowledgments

We appreciate the support from the Health Futures team, especially Roberto Lleras, whose feedback has been crucial to our work.